\documentclass[aps,twocolumn,superscriptaddress]{revtex4}
\usepackage{epsfig}
\usepackage{times}
\usepackage{color}

\begin{document}

\title{Interdependent network reciprocity in evolutionary games}

\author{Zhen Wang}
\affiliation{Department of Physics, Hong Kong Baptist University, Kowloon Tong, Hong Kong}
\affiliation{Center for Nonlinear Studies and the Beijing-Hong Kong-Singapore Joint Center for Nonlinear and Complex Systems, Hong Kong Baptist University, Kowloon Tong, Hong Kong}

\author{Attila Szolnoki}
\affiliation{Institute of Technical Physics and Materials Science, Research Centre for Natural Sciences, Hungarian Academy of Sciences, P.O. Box 49, H-1525 Budapest, Hungary}

\author{Matja{\v z} Perc}
\email{matjaz.perc@uni-mb.si}
\affiliation{Faculty of Natural Sciences and Mathematics, University of Maribor, Koro{\v s}ka cesta 160, SI-2000 Maribor, Slovenia}

\begin{abstract}
Besides the structure of interactions within networks, also the interactions between networks are of the outmost importance. We therefore study the outcome of the public goods game on two interdependent networks that are connected by means of a utility function, which determines how payoffs on both networks jointly influence the success of players in each individual network. We show that an unbiased coupling allows the spontaneous emergence of interdependent network reciprocity, which is capable to maintain healthy levels of public cooperation even in extremely adverse conditions. The mechanism, however, requires simultaneous formation of correlated cooperator clusters on both networks. If this does not emerge or if the coordination process is disturbed, network reciprocity fails, resulting in the total collapse of cooperation. Network interdependence can thus be exploited effectively to promote cooperation past the limits imposed by isolated networks, but only if the coordination between the interdependent networks is not disturbed.
\end{abstract}

\maketitle

Not only are our social interactions limited and thus best described not by well-mixed models but rather by models entailing networks \cite{szabo_pr07, castellano_rmp09, barabasi_np12}, it is also a fact that these networks are often interconnected and indeed very much interdependent \cite{buldyrev_n10, parshani_prl10, huang_xq_pre11, gao_jx_np12}. From the World economy to Google Circles, this interdependence cannot be denied and is all but absent, and in fact it is easily observable. Not necessarily so on the basis of actual links between the networks, which arguably could be the most hidden and difficult to identify feature of such systems, but much more so based on the impact actions in one network will have on the behavior in another network. Several examples are provided in \cite{ball_12}, which demonstrate vividly that we live in a strongly interconnected World. More specifically, seminal works on interdependent networks have shown that even seemingly irrelevant changes in one network can have catastrophic and very much unexpected consequence in another network \cite{buldyrev_n10, zhou_pre12}. Since the evolution of cooperation in human societies also proceeds on such interdependent networks, it is thus of interest to determine to what extent the interdependence influences the outcome of evolutionary games.

Preceding works concerning evolutionary games on individual networks and graphs, as reviewed comprehensively in \cite{szabo_pr07, roca_plr09, perc_bs10}, provide a stimulating starting point for explorations on interdependent networks \cite{wang_z_epl12, gomez-gardenes_srep12, gomez-gardenes_pre12}. The discovery of network reciprocity due to Nowak and May \cite{nowak_n92b}, which demonstrated that the aggregation of cooperators into compact clusters can protect them against invading defectors even in the realm of the most challenging prisoner's dilemma game, launched a spree of research activity aimed at understanding the evolution of cooperation in structured populations. Methods of statistical physics in particular \cite{szabo_pre98}, have proven very valuable for this task, as evidenced by the seminal studies of the evolution of cooperation on small-world \cite{abramson_pre01, kim_bj_pre02}, scale-free \cite{santos_prl05, santos_pnas06}, coevolving \cite{ebel_pre02, zimmermann_pre04}, hierarchical \cite{lee_s_prl11} and bipartite networks \cite{gomez-gardenes_c11}, to name only a few representative examples. Based on these studies it is now established that heterogeneous connections promote cooperation in all main social dilemmas \cite{santos_pnas06}, and that this is a very robust evolutionary outcome \cite{poncela_njp07}, although not immune to the normalization of payoffs \cite{santos_jeb06, masuda_prsb07, tomassini_ijmpc07, szolnoki_pa08} and targeted removal of nodes \cite{perc_njp09}. In fact, heterogeneity in general, not just in terms of players having different degree within a network, has proven to be very effective in maintaining high levels of cooperation in the population \cite{szolnoki_epl07, pinheiro_pone12, ren_pre07, guan_pre07, perc_pre08, santos_n08, perc_njp11, buesser_pre12, santos_jtb12}. Moreover, many coevolutionary rules \cite{perc_bs10} have been introduced that may generate such states spontaneously \cite{pacheco_prl06, fu_pre08, zhang_j_pa11, wu_t_pre09, wu_t_epl09, poncela_njp09, zhang_j_pa10, dai_ql_njp10, lin_yt_pa11}. Adding to this the discovered relevance of the population being structured rather than well-mixed for the effectiveness of reward \cite{szolnoki_epl10, szolnoki_njp12}, punishment \cite{helbing_ploscb10, szolnoki_pre11, perc_njp12}, selection pressure \cite{pinheiro_njp12}, and bargaining \cite{kuperman_epjb08, sinatra_jstat09, iranzo_jtb11, szolnoki_prl12}, it is beyond doubt that the added complexity of the interdependence between networks will lead to the discovery of fascinating new insights concerning how and why we so often choose socially responsible actions over defection, despite the fact that large-scale social experiments suggest otherwise \cite{gracia-lazaro_pnas12}.

\begin{figure}
\centerline{\epsfig{file=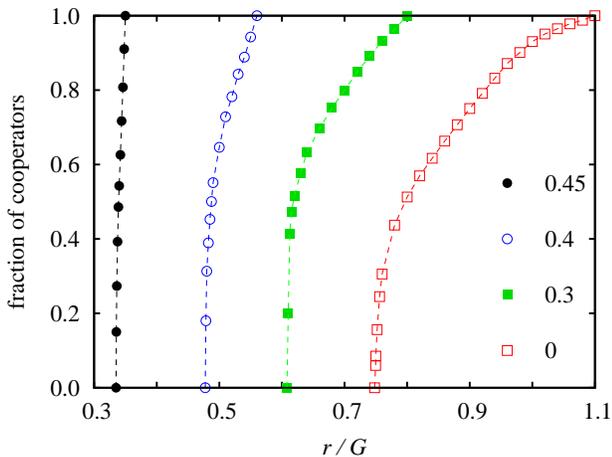,width=8.5cm}}
\caption{\label{motiv} Shifting weight from individual payoffs to average payoffs of nearest neighbors in the evaluation of utility promotes the evolution of public cooperation. Depicted is the stationary fraction of cooperators $\rho_C$ on both lattices in dependence on the normalized multiplication factor $r/G$ for different $\alpha=\beta$ values, as denoted in the figure legend. Due to the payoff-coupling within and between networks, there is a remarkable drop of critical $r/G$ values that are required for cooperators to survive and dominate on both networks. Results were obtained by means of Monte Carlo (MC) simulations using $2 \times L \times L = 2 \times 300 \times 300$ system size and $2 \times 10^5$ MC steps. Averages were determined from the last $10^4$ MC steps.}
\end{figure}

Here we wish to contribute to the subject by studying the evolution of public cooperation on two interdependent networks. In order to focus on the effects that are predominantly due to the interdependence, we consider the square lattice as an elementary interaction network and an evolutionary game that is governed by group interactions, namely the spatial public goods game \cite{szolnoki_pre09c}. The usage of this setup is motivated by our desire to minimize other effects that could stem from network complexity or pairwise interactions. Nevertheless, as we will argue when presenting the results and the newly discovered mechanism of interdependent network reciprocity, our findings should apply to many other social scenarios as well. In general, within the realm of evolutionary games, it is most natural to assume that the interconnectedness will not allow direct strategy exchanges between the networks, but rather affect the utility function of players which influences the strategy imitation probability within a network \cite{wang_z_epl12}. While traditionally the utility is considered to be the payoff of a player as received from the games played with its interaction partners, staging the game on interdependent networks opens up the possibility of alternative, possibly interdependent formulations of utility. A general example is when the payoff of a player is not just due to the games with its neighbors in the current network, but also due to its involvement in another network. Two individuals with strong ties but working in separate environments, or the relation between two companies of the same group but in different countries, can be characterized this way. We make use of these possibilities and propose a simple network-symmetric formulation of utility based on the payoff of player in the first network, as well as the average payoffs of its neighbors on both the original and the second network. While this formulation is not intended to model any specific scenario, it captures relevantly the essence of interdependent utilities. Notably, considering the payoffs of neighbors provides a natural coupling within a network, which can then be compared with the consequence of coupling between networks. Importantly, as the definition of utility is symmetric with respect to the two networks, there is no master-slave relation between them. As we will show, this setup leads us to the observation of interdependent network reciprocity. In particular, if related players on both networks aggregate into compact cooperative domains, then this form of reciprocity is much more powerful than traditional network reciprocity on an isolated network. On the other hand, even if some cooperators aggregate into a compact domain in one network, they ultimately fail to survive in the absence of similarly aggregated partners in the other network. Hence the term interdependent network reciprocity.

\section*{Results}

We first present results for $\alpha=\beta$ (see Methods for the description of the model and parameters), in which case the focus is mainly on the transition of importance from individual payoffs to average payoffs of nearest neighbors in the evaluation of utility. Naturally, as $\beta$ increases the interdependence between the two networks grows too (see Eq.~\ref{utility}), but it is impossible to attribute the reported effects unequivocally either to the growing influence of group payoffs or to the growing interdependence. As results presented in Fig.~\ref{motiv} show, simultaneously increasing both $\alpha$ and $\beta$ continuously lowers the minimally required multiplication factor $r$ that is needed for cooperator to survive. What is more, the span of $r$ where cooperators and defectors coexist shrinks too, ultimately leading also to an ever-faster arrival to the pure $C$ phase as $r$ increases further. More precisely, the minimally required $r/G$ drops from $0.748$ for $\alpha=\beta=0$ to $r/G=0.33$ for $\alpha=\beta=0.45$, and correspondingly, the required $r/G$ for complete cooperator dominance drops from $r/G=1.09$ to just $r/G=0.35$. This is certainly impressive and motivating for wanting to understand the mechanism behind the remarkable promotion of public cooperation. Even before further results are presented, however, it is clear that the transition of weight from individual payoffs to average payoffs of nearest neighbors in determining the utility of each player plays an important role. In particular, the transition magnifies the benefit of $C-C$ coupling on both networks, yet this effect alone would be unable to improve the viability of cooperators to such an extent as reported in Fig.~\ref{motiv}.

\begin{figure}[b]
\centerline{\epsfig{file=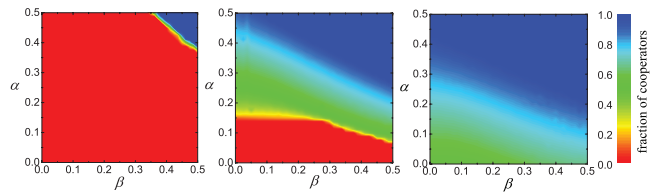,width=8.5cm}}
\caption{\label{separate} Shifting weight from individual payoffs to average payoffs of nearest neighbors promotes the evolution of cooperation, but it is important whether the shift entails the average payoffs of neighbors on the host network ($\alpha$) or the average payoffs of neighbors on the other, interdependent network ($\beta$). Depicted are contour plots showing the stationary fraction of cooperators $\rho_C$ in dependence on both $\alpha$ and $\beta$ for $r/G=0.4$ (left), $0.7$ (middle), and 0.8 (right). Increasing $\alpha$ at any given $\beta$ promotes the evolution of cooperation. The same is true vice versa, only that the positive effect depends much more on $\alpha$ and $r/G$, thus highlight the importance of an intact network reciprocity on each individual network for increasing $\beta$ to work. Details of Monte Carlo simulations are the same as described in Fig.~\ref{motiv}.}
\end{figure}

It is therefore of interest to decouple the impact of $\alpha$ and $\beta$ by eliminating the $\alpha=\beta$ restrain, as presented in Fig.~\ref{separate}, where contour plots encode the stationary fraction of cooperators $\rho_C$ for three different values of $r/G$. Regardless of the multiplication factor, it can be observed that at any given value of $\beta$, increasing $\alpha$ will elevate $\rho_C$. Note that increasing $\alpha$ simply elevates the relevance of the payoffs of nearest neighbors on the expense of individual payoffs on any given network, without modifying the level of interdependence between them. The impact of $\beta$ at any given $\alpha$, however, is less clear-cut. Note that increasing $\beta$, however, also elevates the relevance of the payoffs of nearest neighbors on the expense of individual payoffs, but on separate networks, thus increasing the level of interdependence between them. In general, increasing $\beta$ at a fixed value of $\alpha$ also elevates $\rho_C$ levels, thus suggesting that an increase in the interdependence of the utility function also promotes public cooperation, similarly as the increase of $\alpha$ at any given $\beta$. Yet the efficiency of increasing $\beta$ depends much more on $\alpha$ than vice versa. For low values of $\alpha$ increasing $\beta$ has a much more marginal impact than if increasing $\beta$ at high values of $\alpha$. This difference in impact also grows as $r/G$ decreases. Both these facts indicate that network reciprocity on each individual network is crucial for higher $\beta$ to take effect, as both lower values of $\alpha$ and $r/G$ make it increasingly difficult for cooperators to form compact domains.

\begin{figure}
\centerline{\epsfig{file=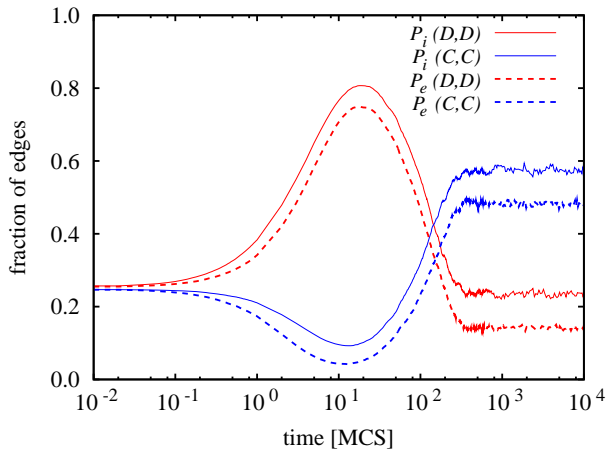,width=8.32cm}}
\caption{\label{pairs} Cooperators and defectors on the two interdependent networks share exactly the same fate during the course of evolution, i.e., as soon as the density of cooperator (defector) pairs starts decreasing (increasing) on one network, the same happens on the other network. In the legend $P_i$ and $P_e$ mark the corresponding pair configuration probabilities within (internal) and between (external) networks, respectively. The evolution is completely synchronous, with quantitative differences emerging only due to the applied relatively small value of $\beta$. Notably, for $\beta \to 0.5$ the difference would be virtually none. Parameter values used were: $r/G=0.7$, $\alpha=0.15$, and $\beta=0.5$, at $L=300$.}
\end{figure}

Having then established the relevance of traditional network reciprocity on each individual network, it is next instructive to monitor the evolution of cooperation separately on the two networks. In order to do so, we first examine how the fraction of $C-C$ and $D-D$ pairs varies separately on network A and B for a characteristic combination of game parameters, as presented in Fig.~\ref{pairs}. Here $P_i$ and $P_e$ denote the fraction of the related pair configurations within and between networks, respectively. Accordingly, $P_i(C,C)$ is the probability to find a $C-C$ pair in any given network, while $P_e(C,C)$ is the probability to find a $C-C$ link between the two networks. Besides the initial fall of cooperators before the eventual rise, which is the hallmark signature indicating the spontaneous emergence of spatial reciprocity, it is important to note just how synchronized the rise (fall) and fall (rise) of $D-D$ ($C-C$) pairs on the two networks is. There is only a small quantitative difference inferable, but otherwise both cooperators and defectors on the two networks share completely the same fate.

\begin{figure}
\centerline{\epsfig{file=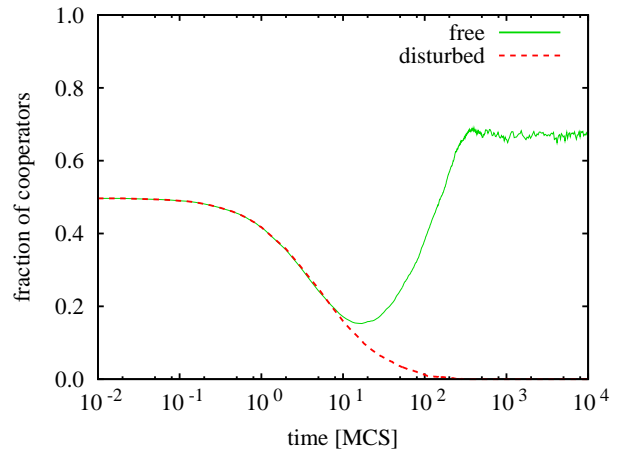,width=8.5cm}}
\caption{\label{disturb} Mixing the distribution of players in one of the two networks disrupts the synchronous evolution of the two strategies, which remarkably has dire consequences for the fate of cooperation on the other network, despite the fact that the coupling of payoffs within the network is still significant. The evolution of cooperation in the latter, measured by the density of cooperators, is denoted dashed red. For comparison, solid green line shows the time dependence in the same network while the strategy evolution is not disturbed in the other network. This experiment highlights that while traditional network reciprocity alone can be inefficient, the chances of cooperators improve dramatically if interdependent network reciprocity is allowed to take effect. The applied system size was $2 \times 200 \times 200$ at $r/G=0.7$, $\alpha=0.15$, and $\beta=0.5$.}
\end{figure}

The question that now poses itself is just how important the synchrony of the evolutionary process on the two interdependent networks is. As evidenced by the results presented in Fig.~\ref{disturb}, the importance can hardly be overstated. If we prevent the spontaneous formation of cooperative domains in one of the networks by applying a permanent mixing of strategies, then this will negatively influence the evolution of cooperation in the other (undisturbed) network as well. On the latter, the cooperators will die out despite the fact that the coupling within the network is still functioning via nonzero $\alpha$, and that thus the evolution of cooperation ought to be supported on this undisturbed network. As can be observed, this intervention on one of the networks will not only prevent the emergence of synchronized evolution on the two networks, but will also result in full defection in the undisturbed network in spite of the fact that the two networks are not physically connected (the corresponding evolution in the unperturbed network is denoted by ``disturbed'' in Fig.~\ref{disturb}). For comparison, we also plot the evolution in the same network when the emergence of parallel evolution was not disturbed (denoted by ``free'' in Fig.~\ref{disturb}). As the ``free'' line shows, cooperators can eventually spread widely even at parameter values where traditional single-network reciprocity would fail completely.

\begin{figure}
\centerline{\epsfig{file=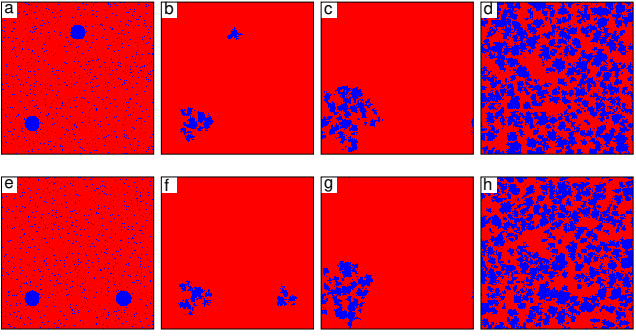,width=8.5cm}}
\caption{\label{snaps} Direct observation of interdependent network reciprocity. From left to right, panels (a)-(d) depict the evolution of cooperation in one network, while panels (e)-(h) show the same for the other network. Starting from a prepared initial state, only those circular cooperative domains that are initially present on both networks, and which can therefore make immediate use of interdependent network reciprocity, survive and eventually spread across both networks. Cooperators that are initially distributed uniformly at random, as well as clustered cooperators present on one but not the other network, surrender under the evolutionary pressure from defectors already at the early stages of evolution. The applied system size was $2 \times 200 \times 200$ at $r/G=0.66$, $\alpha=0.15$, and $\beta=0.5$. Snapshots were taken at $t=0$, $100$, $500$ and $10000$ MC steps.}
\end{figure}

These results lead us to the coining of the term ``interdependent network reciprocity'', simply because if network reciprocity fails on one network, it will inevitably fail also on the other network. If it remains intact on both, however, and if the interdependence via the utility function is sufficiently strong to initialize spontaneous coordination between the two networks, then interdependent network reciprocity will strongly outperform traditional single-network reciprocity, in fact leading to much higher levels of public cooperation, as evidenced in Figs.~\ref{motiv} and \ref{separate}. The mechanism of interdependent network reciprocity can be demonstrated most beautifully by means of characteristic snapshots of the two interdependent networks as obtained from a prepared initial state. Figure~\ref{snaps} shows that only the one circular cooperative domain is able to survive the initial onslaught of defectors, namely the one that is initially present on both networks, but not the other two which are initially present on only one network. Notably, other cooperators that are initially distributed uniformly at random, and which could potentially be saved by traditional single-network reciprocity, die out fast too. Thus, only those cooperators that can make immediate use of interdependent network reciprocity survive and eventually spread across both networks.

\begin{figure}
\centerline{\epsfig{file=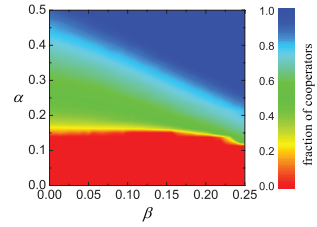,width=4cm}}
\caption{\label{general} Interdependent network reciprocity works also one more than two symmetrically interdependent networks. Like for two networks, for three interdependent networks shifting weight from individual payoffs to average payoffs of nearest neighbors also promotes the evolution of cooperation, but it is important whether the shift entails the average payoffs of neighbors on the host network ($\alpha$) or the average payoffs of neighbors on the other, interdependent network ($\beta$). For interdependent network reciprocity to work best, a threshold value of $\alpha \approx 0.17$ needs to be surpassed. The applied system size was $3 \times L \times L = 3 \times 400 \times 400$ and $5 \times 10^5$ MC steps were used, while the value of $r/G=0.7$ was the same as used in the middle panel of Fig.~\ref{separate}, which thus also serves as a comparison reference.}
\end{figure}

Lastly it is of interest to elaborate on the generality of the identified interdependent network reciprocity. To do so, we consider an alternative setup entailing three symmetrically interdependent networks with the governing utility function being
\begin{equation}
U_X= (1-\alpha-2\beta) P_x + \alpha \langle P_x \rangle + \beta \langle P_{x^\prime} \rangle + \beta \langle P_{x^{\prime \prime}} \rangle \,.
\label{three}
\end{equation}
Here $x^{\prime}$ and $x^{\prime \prime}$ denote the related players of player $x$ in the corresponding networks. In this way every network is connected with two other networks in a symmetric way, without any master-slave preferences. As in the originally studied model, strategy imitation is allowed only between players residing on the same network, but never between players from different networks. As results presented in Fig.~\ref{general} show, the impact of different $\alpha$ and $\beta$ values is qualitatively the same as reported in Fig.~\ref{separate} for the two-network setup. In particular, it can be observed that at any given value of $\beta$, increasing $\alpha$ will elevate $\rho_C$, while increasing $\beta$ at a fixed value of $\alpha$ also elevates $\rho_C$ levels, but it requires a threshold value of $\alpha$ to be surpassed for the interdependent network reciprocity to become really effective. Altogether, these results indicate clearly that the spontaneous emergence of coordination and subsequent synchronization of the evolutionary process is possible also if more than two networks are interdependent. Further on the generality of interdependent network reciprocity, we remind that the outcome of public goods games is in general independent of the interaction topology \cite{szolnoki_pre09c}, and therefore a similar impact can be expected also if the topology of individual networks is different from the square lattice. Moreover, since social dilemma games governed by pairwise interactions were recently also found to be susceptible to the ``wisdom of group'' effect \cite{szolnoki_srep12}, we argue that interdependent network reciprocity should be observable straightforwardly by them as well. More challenging setups, which would be of interest to study more accurately in the future, would entail determining to what degree the network topology of each individual network forming the interdependent system is allowed to differ before for interdependent network reciprocity to remain intact.

\section*{Discussion}
Summarizing, we have studied the evolution of cooperation in the public goods game on interdependent networks that are subject to interconnectedness by means of a network-symmetric definition of utility. Strategy imitation has been allowed only between players residing on the same network, but not between players on different networks. We have shown first that, in general, increasing the relevance of the average payoff of nearest neighbors on the expense of individual payoffs in the evaluation of utility increases the survivability of cooperators. More importantly, however, we have shown that while doing this for the coupling between individual and group payoffs on a given network simply reinforces network reciprocity, applying the same coupling of payoffs between networks has a much more interesting and in fact unexpected outcome. In particular, we have identified the spontaneous emergence of the so-called interdependent network reciprocity. The later implies that if cooperative players on network A aggregate into compact domains while the corresponding players on network B fail to do so (or vice versa due to the network-symmetric definition of utility), the mechanism of network reciprocity will ultimately fail on both networks. On the other hand, if players on network A aggregate and the corresponding players on network B do too, then the evolution of cooperation will be promoted much stronger than based on traditional single-network reciprocity alone. At this point it is important to note that players between the networks are not physically connected, nor are they allowed to imitate strategy from one another. The coordination of clustering of cooperators on both networks is a spontaneous process that is thus brought about solely by the interdependence of utility. We have shown the mechanism of interdependent network reciprocity to be robust against the number of interdependent networks, and we have provided arguments in favor of robustness against variations of the structure of each individual network, as well against the type of the governing evolutionary game played. Overall, our research indicates that the interdependence of interaction networks can be exploited effectively to promote public cooperation past the limits imposed by isolated networks, although this is subject to warranting undisturbed coordination of the distribution of strategies on the two networks, which in turn requires a sufficiently strong interdependence of utility.\\

\section*{Methods}
The public goods game on both networks is staged on a $L \times L$ square lattice with periodic boundary conditions, where players are arranged into overlapping groups of size $G=5$. Every player is thus surrounded by its $k=G-1$ nearest neighbors and is a member in $g=G$ different groups. Initially each player on site $x$ in network A and on site $x^\prime$ in network B is designated either as a cooperator $C$ or defector $D$ with equal probability. The accumulation of payoffs $P_x$ and $P_{x^\prime}$ on both networks follows the same standard procedure. Namely, in each group cooperators contribute $1$ to the public good while defectors contribute nothing. The sum of contributions is subsequently multiplied by the factor $r>1$, reflecting the synergetic effects of cooperation, and the resulting amount is equally shared amongst the $G$ group members. In each group the payoff obtained is $P_x^g$ on network A and $P_{x^\prime}^g$ on network B, while the total amount received in all the groups is thus $P_x = \sum_g P_x^g$ and $P_{x^\prime} = \sum_g P_{x^\prime}^g$.

While the two networks are not physically connected, interdependence is introduced via the utility function
\begin{equation}
U_x = (1-\alpha-\beta) P_x + \alpha \langle P_x \rangle + \beta \langle P_{x^\prime} \rangle\,,
\label{utility}
\end{equation}
where $\langle P_x \rangle$ and $\langle P_{x^\prime} \rangle$ are the average payoffs of all four neighbors of players $x$ and $x^\prime$ on their host lattices, respectively. It is important to note that this definition is network-symmetric. The evaluation of $U_{x^\prime}$ is identical if the indexes ($^\prime$) for the two networks are switched, and thus there is no master-slave relation between them. The later was considered previously in \cite{wang_z_epl12}, but it constitutes and entirely different setup. In accordance with the proposed symmetric definition of utility as given by Eq.~\ref{utility}, increasing $\alpha$ and/or $\beta$, within the limitations of the condition $(\alpha+\beta)<1$, puts more emphasis on the payoffs of the neighbors, while at the same time decreasing the relevance of individual payoffs. As expected, larger values of $\alpha$ promote cooperation, but the consequence of increasing $\beta$ is hardly foreseeable as there is no direct information exchange between the networks via strategy imitation. Setting $\beta=0$ decouples the two networks and more specifically, at $\alpha=\beta=0$ we regain the traditional spatial public goods game on both square lattices, where, as reported in \cite{szolnoki_pre09c}, cooperators die out if the multiplication factor $r$ decreases below a threshold, depending also on the uncertainty governing the strategy adoptions $K$ (see Eq.~\ref{fermi} below). For $K=0.5$, which we will use throughout this work without loss of generality, the critical value is $r/G=0.748$.

Following the determination of utilities of players according to Eq.~\ref{utility}, strategy imitation is possible only between nearest neighbors on any given lattice, but never between players residing on different networks. We emphasize that this restrain, in addition to the network-symmetric formulation of utility, are the two key considerations of the present setup. Accordingly, on network A player $x$ can adopt the strategy $s_{y}$ of one of its randomly chosen nearest neighbors $y$ with a probability determined by the Fermi function
\begin{equation}
W(s_{y} \rightarrow s_{x})=\frac{1}{1+\exp[(U_x-U_y)/K]} \,,
\label{fermi}
\end{equation}
where the utility $U_{y}$ of player $y$ is evaluated identically as for player $x$. The probability of strategy invasion from player $y^\prime$ to player $x^\prime$ on network B is determined likewise, only that utilities $U_x^\prime$ and $U_y^\prime$ are used. Simulations of the model were performed by means of a random sequential update, where each player on both networks had a chance to pass its strategy once on average during a Monte Carlo (MC) step. The linear system size was varied from $L=100$ to $600$ in order to avoid finite size effects, and the equilibration required up to $10^5$ MC steps.

\begin{acknowledgments}
This research was supported by the Hungarian National Research Fund (Grant K-101490) and the Slovenian Research Agency (Grant J1-4055).
\end{acknowledgments}


\begin{thebibliography}{10}

\bibitem{szabo_pr07}
Szab{\'o}, G. and F{\'a}th, G.
\newblock Evolutionary games on graphs.
\newblock {\em Phys. Rep.}{ \bf 446}, 97--216 (2007).

\bibitem{castellano_rmp09}
Castellano, C., Fortunato, S., and Loreto, V.
\newblock Statistical physics of social dynamics.
\newblock {\em Rev. Mod. Phys.}{ \bf 81}, 591--646 (2009).

\bibitem{barabasi_np12}
Barab{\'a}si, A.-L.
\newblock The network takeover.
\newblock {\em Nature Physics}{ \bf 8}, 14--16 (2012).

\bibitem{buldyrev_n10}
Buldyrev, S.~V., Parshani, R., Paul, G., Stanley, H.~E., and Havlin, S.
\newblock Catastrophic cascade of failures in interdependent networks.
\newblock {\em Nature}{ \bf 464}, 1025--1028 (2010).

\bibitem{zhou_pre12}
Zhou, D., Stanley, H.~E., D'Agostino, G., and Scala, A.
\newblock Assortativity decreases the robustness of interdependent networks.
\newblock {\em Phys. Rev. E}{ \bf 86},  066103 (2012).

\bibitem{parshani_prl10}
Parshani, R., Buldyrev, S.~V., and Havlin, S.
\newblock Interdependent networks: Reducing the coupling strength leads to a
  change from a first to second order percolation transition.
\newblock {\em Phys. Rev. Lett.}{ \bf 105}, 048701 (2010).

\bibitem{huang_xq_pre11}
Huang, X., Gao, J., Buldyrev, S.~V., Havlin, S., and Stanley, H.~E.
\newblock Robustness of interdependent networks under targeted attack.
\newblock {\em Phys. Rev. E}{ \bf 83}, 065101(R) (2011).

\bibitem{gao_jx_np12}
Gao, J., Buldyrev, S.~V., Stanley, H.~E., and Havlin, S.
\newblock Networks formed from interdependent networks.
\newblock {\em Nature Physics}{ \bf 8}, 40--48 (2012).

\bibitem{ball_12}
Ball, P.
\newblock {\em Why Society is a Complex Matter}.
\newblock Springer, Berlin Heidelberg,  (2012).

\bibitem{roca_plr09}
Roca, C.~P., Cuesta, J.~A., and S{\'a}nchez, A.
\newblock Evolutionary game theory: Temporal and spatial effects beyond
  replicator dynamics.
\newblock {\em Phys. Life Rev.}{ \bf 6}, 208--249 (2009).

\bibitem{perc_bs10}
Perc, M. and Szolnoki, A.
\newblock Coevolutionary games -- a mini review.
\newblock {\em BioSystems}{ \bf 99}, 109--125 (2010).

\bibitem{wang_z_epl12}
Wang, Z., Szolnoki, A., and Perc, M.
\newblock Evolution of public cooperation on interdependent networks: The
  impact of biased utility functions.
\newblock {\em EPL}{ \bf 97}, 48001 (2012).

\bibitem{gomez-gardenes_srep12}
G{\'o}mez-Garde{\~n}es, J., Reinares, I., Arenas, A., and Flor{\' \i}a, L.~M.
\newblock Evolution of cooperation in multiplex networks.
\newblock {\em Sci. Rep.}{ \bf 2}, 620 (2012).

\bibitem{gomez-gardenes_pre12}
G{\'o}mez-Garde{\~n}es, J., Gracia-L{\'a}zaro, C., Flor{\'i}a, L.~M., and
  Moreno, Y.
\newblock Evolutionary dynamics on interdependent populations.
\newblock {\em Phys. Rev. E}{ \bf 86}, 056113 (2012).

\bibitem{nowak_n92b}
Nowak, M.~A. and May, R.~M.
\newblock Evolutionary games and spatial chaos.
\newblock {\em Nature}{ \bf 359}, 826--829 (1992).

\bibitem{szabo_pre98}
Szab{\'o}, G. and T{\H{o}}ke, C.
\newblock Evolutionary prisoner's dilemma game on a square lattice.
\newblock {\em Phys. Rev. E}{ \bf 58}, 69--73 (1998).

\bibitem{abramson_pre01}
Abramson, G. and Kuperman, M.
\newblock Social games in a social network.
\newblock {\em Phys. Rev. E}{ \bf 63}, 030901(R) (2001).

\bibitem{kim_bj_pre02}
Kim, B.~J., Trusina, A., Holme, P., Minnhagen, P., Chung, J.~S., and Choi,
  M.~Y.
\newblock Dynamic instabilities induced by asymmetric influence: Prisoner's
  dilemma game in small-world networks.
\newblock {\em Phys. Rev. E}{ \bf 66}, 021907 (2002).

\bibitem{santos_prl05}
Santos, F.~C. and Pacheco, J.~M.
\newblock Scale-free networks provide a unifying framework for the emergence of
  cooperation.
\newblock {\em Phys. Rev. Lett.}{ \bf 95}, 098104 (2005).

\bibitem{santos_pnas06}
Santos, F.~C., Pacheco, J.~M., and Lenaerts, T.
\newblock Evolutionary dynamics of social dilemmas in structured heterogeneous
  populations.
\newblock {\em Proc. Natl. Acad. Sci. USA}{ \bf 103}, 3490--3494 (2006).

\bibitem{ebel_pre02}
Ebel, H. and Bornholdt, S.
\newblock Coevolutionary games on networks.
\newblock {\em Phys. Rev. E}{ \bf 66}, 056118 (2002).

\bibitem{zimmermann_pre04}
Zimmermann, M.~G., Egu{\'{\i}}luz, V., and Miguel, M.~S.
\newblock Coevolution of dynamical states and interactions in dynamic networks.
\newblock {\em Phys. Rev. E}{ \bf 69}, 065102(R) (2004).

\bibitem{lee_s_prl11}
Lee, S., Holme, P., and Wu, Z.-X.
\newblock Emergent hierarchical structures in multiadaptive games.
\newblock {\em Phys. Rev. Lett.}{ \bf 106}, 028702 (2011).

\bibitem{gomez-gardenes_c11}
G{\'o}mez-Garde{\~n}es, J., Romance, M., Criado, R., Vilone, D., and
  S{\'a}nchez, A.
\newblock Evolutionary games defined at the network mesoscale: The public goods
  game.
\newblock {\em Chaos}{ \bf 21}, 016113 (2011).

\bibitem{poncela_njp07}
Poncela, J., G{\'o}mez-Garde{\~n}es, J., Flor{\' \i}a, L.~M., and Moreno, Y.
\newblock Robustness of cooperation in the evolutionary prisoner's dilemma on
  complex systems.
\newblock {\em New J. Phys.}{ \bf 9}, 184 (2007).

\bibitem{santos_jeb06}
Santos, F.~C. and Pacheco, J.~M.
\newblock A new route to the evolution of cooperation.
\newblock {\em J. Evol. Biol.}{ \bf 19}, 726--733 (2006).

\bibitem{masuda_prsb07}
Masuda, N.
\newblock Participation costs dismiss the advantage of heterogeneous networks
  in evolution of cooperation.
\newblock {\em Proc. R. Soc. B}{ \bf 274}, 1815--1821 (2007).

\bibitem{tomassini_ijmpc07}
Tomassini, M., Luthi, L., and Pestelacci, E.
\newblock Social dilemmas and cooperation in complex networks.
\newblock {\em Int. J. Mod. Phys. C}{ \bf 18}, 1173--1185 (2007).

\bibitem{szolnoki_pa08}
Szolnoki, A., Perc, M., and Danku, Z.
\newblock Towards effective payoffs in the prisoner's dilemma game on
  scale-free networks.
\newblock {\em Physica A}{ \bf 387}, 2075--2082 (2008).

\bibitem{perc_njp09}
Perc, M.
\newblock Evolution of cooperation on scale-free networks subject to error and
  attack.
\newblock {\em New J. Phys.}{ \bf 11}, 033027 (2009).

\bibitem{szolnoki_epl07}
Szolnoki, A. and Szab{\'o}, G.
\newblock Cooperation enhanced by inhomogeneous activity of teaching for
  evolutionary prisoner's dilemma games.
\newblock {\em EPL}{ \bf 77}, 30004 (2007).

\bibitem{pinheiro_pone12}
Pinheiro, F.L., Pacheco, J.M., and Santos, F.C.
\newblock From Local to Global Dilemmas in Social Networks.
\newblock {\em PLoS ONE}{ \bf 7}, e32114 (2012).

\bibitem{ren_pre07}
Ren, J., Wang, W.-X., and Qi, F.
\newblock Randomness enhances cooperation: coherence resonance in evolutionary
  game.
\newblock {\em Phys. Rev. E}{ \bf 75}, 045101(R) (2007).

\bibitem{guan_pre07}
Guan, J.-Y., Wu, Z.-X., and Wang, Y.-H.
\newblock Effects of inhomogeneous activity of players and noise on cooperation
  in spatial public goods games.
\newblock {\em Phys. Rev. E}{ \bf 76}, 056101 (2007).

\bibitem{perc_pre08}
Perc, M. and Szolnoki, A.
\newblock Social diversity and promotion of cooperation in the spatial
  prisoner's dilemma game.
\newblock {\em Phys. Rev. E}{ \bf 77}, 011904 (2008).

\bibitem{santos_n08}
Santos, F.~C., Santos, M.~D., and Pacheco, J.~M.
\newblock Social diversity promotes the emergence of cooperation in public
  goods games.
\newblock {\em Nature}{ \bf 454}, 213--216 (2008).

\bibitem{perc_njp11}
Perc, M.
\newblock Does strong heterogeneity promote cooperation by group interactions?
\newblock {\em New J. Phys.}{ \bf 13}, 123027 (2011).

\bibitem{buesser_pre12}
Buesser, P. and Tomassini M.
\newblock Evolution of cooperation on spatially embedded networks.
\newblock {\em Phys. Rev. E}{ \bf 86}, 066107 (2012).

\bibitem{santos_jtb12}
Santos, F.~C., Pinheiro, F., Lenaerts, T., and Pacheco, J.~M.
\newblock Role of diversity in the evolution of cooperation.
\newblock {\em J. Theor. Biol.}{ \bf 299}, 88--96 (2012).

\bibitem{pacheco_prl06}
Pacheco, J.~M., Traulsen, A., and Nowak, M.~A.
\newblock Coevolution of strategy and structure in complex networks with
  dynamical linking.
\newblock {\em Phys. Rev. Lett.}{ \bf 97}, 258103 (2006).

\bibitem{fu_pre08}
Fu, F. and Wang, L.
\newblock Coevolutionary dynamics of opinions and networks: From diversity to
  uniformity.
\newblock {\em Phys. Rev. E}{ \bf 78}, 016104 (2008).

\bibitem{zhang_j_pa11}
Zhang, J., Wang, W.-Y., Du, W.-B., and Cao, X.-B.
\newblock Evolution of cooperation among mobile agents with heterogenous view
  radius.
\newblock {\em Physica A}{ \bf 390}, 2251--2257 (2011).

\bibitem{wu_t_pre09}
Wu, T., Fu, F., and Wang, L.
\newblock Partner selections in public goods games with constant group size.
\newblock {\em Phys. Rev. E}{ \bf 80}, 026121 (2009).

\bibitem{wu_t_epl09}
Wu, T., Fu, F., and Wang, L.
\newblock Individual's expulsion to nasty environment promotes cooperation in
  public goods games.
\newblock {\em EPL}{ \bf 88}, 30011 (2009).

\bibitem{poncela_njp09}
Poncela, J., G{\'o}mez-Garde{\~n}es, J., Traulsen, A., and Moreno, Y.
\newblock Evolutionary game dynamics in a growing structured population.
\newblock {\em New J. Phys.}{ \bf 11}, 083031 (2009).

\bibitem{zhang_j_pa10}
Zhang, J., Cao, X.-B., Du, W.-B., and Cai, K.-Q.
\newblock Evolution of chinese airport network.
\newblock {\em Physica A}{ \bf 389}, 3922--3931 (2010).

\bibitem{dai_ql_njp10}
Dai, Q., Li, H., Cheng, H., Li, Y., and Yang, J.
\newblock Double-dealing behavior potentially promotes cooperation in
  evolutionary prisoner's dilemma games.
\newblock {\em New J. Phys.}{ \bf 12}, 113015 (2010).

\bibitem{lin_yt_pa11}
Lin, Y.-T., Yang, H.-X., Wu, Z.-X., and Wang, B.-H.
\newblock Promotion of cooperation by aspiration-induced migration.
\newblock {\em Physica A}{ \bf 390}, 77--82 (2011).

\bibitem{szolnoki_epl10}
Szolnoki, A. and Perc, M.
\newblock Reward and cooperation in the spatial public goods game.
\newblock {\em EPL}{ \bf 92}, 38003 (2010).

\bibitem{szolnoki_njp12}
Szolnoki, A. and Perc, M.
\newblock Evolutionary advantages of adaptive rewarding.
\newblock {\em New J. Phys.}{ \bf 14}, 093016 (2012).

\bibitem{helbing_ploscb10}
Helbing, D., Szolnoki, A., Perc, M., and Szab{\'o}, G.
\newblock Evolutionary establishment of moral and double moral standards
  through spatial interactions.
\newblock {\em PLoS Comput. Biol.}{ \bf 6}, e1000758 (2010).

\bibitem{szolnoki_pre11}
Szolnoki, A., Szab{\'o}, G., and Perc, M.
\newblock Phase diagrams for the spatial public goods game with pool
  punishment.
\newblock {\em Phys. Rev. E}{ \bf 83}, 036101 (2011).

\bibitem{perc_njp12}
Perc, M. and Szolnoki, A.
\newblock Self-organization of punishment in structured populations.
\newblock {\em New J. Phys.}{ \bf 14}, 043013 (2012).

\bibitem{pinheiro_njp12}
Pinheiro, F.~L., Santos, F.~C., and Pacheco, J.~M.
\newblock How selection pressure changes the nature of social dilemmas in structured populations.
\newblock {\em New J. Phys.}{ \bf 14}, 073035 (2012).

\bibitem{kuperman_epjb08}
Kuperman, M.~N. and Risau-Gusman, S.
\newblock The effect of topology on the spatial ultimatum game.
\newblock {\em Eur. Phys. J. B}{ \bf 62}, 233--238 (2008).

\bibitem{sinatra_jstat09}
Sinatra, R., Iranzo, J., G{\'o}mez-Garde{\~n}es, J., Flor\'{\i}a, L.~M.,
  Latora, V., and Moreno, Y.
\newblock The ultimatum game in complex networks.
\newblock {\em J. Stat. Mech.}{ \bf }, P09012 (2009).

\bibitem{iranzo_jtb11}
Iranzo, J., Rom{\'a}n, J., and S{\'a}nchez, A.
\newblock The spatial ultimatum game revisited.
\newblock {\em J. Theor. Biol.}{ \bf 278}, 1--10 (2011).

\bibitem{szolnoki_prl12}
Szolnoki, A., Perc, M., and Szab{\'o}, G.
\newblock Defense mechanisms of empathetic players in the spatial ultimatum
  game.
\newblock {\em Phys. Rev. Lett.}{ \bf 109}, 078701 (2012).

\bibitem{gracia-lazaro_pnas12}
Gracia-L{\'a}zaro, C., Ferrer, A., Ruiz, G., Taranc{\'o}n, A., Cuesta, J.,
  S{\'a}nchez, A., and Moreno, Y.
\newblock Heterogeneous networks do not promote cooperation when humans play a
  prisoner's dilemma.
\newblock {\em Proc. Natl. Acad. Sci. USA}{ \bf 109}, 12922--12926 (2012).

\bibitem{szolnoki_pre09c}
Szolnoki, A., Perc, M., and Szab{\'o}, G.
\newblock Topology-independent impact of noise on cooperation in spatial public
  goods games.
\newblock {\em Phys. Rev. E}{ \bf 80}, 056109 (2009).

\bibitem{szolnoki_srep12}
Szolnoki, A., Wang, Z., and Perc, M.
\newblock Wisdom of groups promotes cooperation in evolutionary social
  dilemmas.
\newblock {\em Sci. Rep.}{ \bf 2}, 576 (2012).

\end{thebibliography}
\end{document}